\documentclass[letter]{aa} 

\usepackage{graphicx}
\usepackage{txfonts}
\usepackage[normalem]{ulem}
\usepackage{color}
\usepackage{esvect}

\definecolor{purple}{rgb}{0.5,0,0.87}

\usepackage[colorlinks,allcolors=blue]{hyperref}

\begin{document}

   \title{Super-massive black hole wake or bulgeless edge-on galaxy?}

   \author{Jorge S\'anchez Almeida
          \inst{1,2}
          \and Mireia Montes\inst{1,2}
          \and
          Ignacio Trujillo\inst{1,2}
          }

\institute{Instituto de Astrof\'\i sica de Canarias, c/ Vía Láctea s/n, E-38205, La Laguna, Tenerife,  Spain
      %
      \and
      Departamento de Astrof\'\i sica, Universidad de La Laguna, E-38203, La Laguna, Tenerife, Spain
      }

   \date{Received \today; accepted \dots }

 
  \abstract{  
  \citet{2023arXiv230204888V} reported the serendipitous discovery of a thin linear object interpreted as the trail of star-forming regions left behind by a runaway supermassive black hole (SMBH) kicked out from the center of a galaxy. Despite the undeniable interest in the idea, the actual physical interpretation is not devoid of difficulty.  The wake of a SMBH produces only small perturbations on the external medium, which has to be in exceptional physical conditions to collapse gravitationally and form a long (40\,kpc) massive ($3\times 10^{9}\,{\rm M_\odot}$) stellar trace in only 39\,Myr. Here we offer a more conventional explanation: the stellar trail is a bulgeless galaxy viewed edge-on. This interpretation is supported by the fact that its position--velocity curve resembles a rotation curve which, together with its stellar mass, puts the object right on top of the Tully-Fisher relation characteristic of disk galaxies. Moreover, the rotation curve ($V_{max}\sim$$110\, {\rm km\,s^{-1}}$), stellar mass, extension, width ($z_0\sim$1.2\,kpc), and surface brightness profile of the object are very much like those of IC\,5249, a well-known local bulgeless edge-on galaxy.  
  These observational facts are difficult to interpret within the SMBH wake scenario. We discuss in detail the pros and cons of the two options. 
}

   \maketitle

\begin{figure}[!]
   \centering
   \includegraphics[width=1.\linewidth]{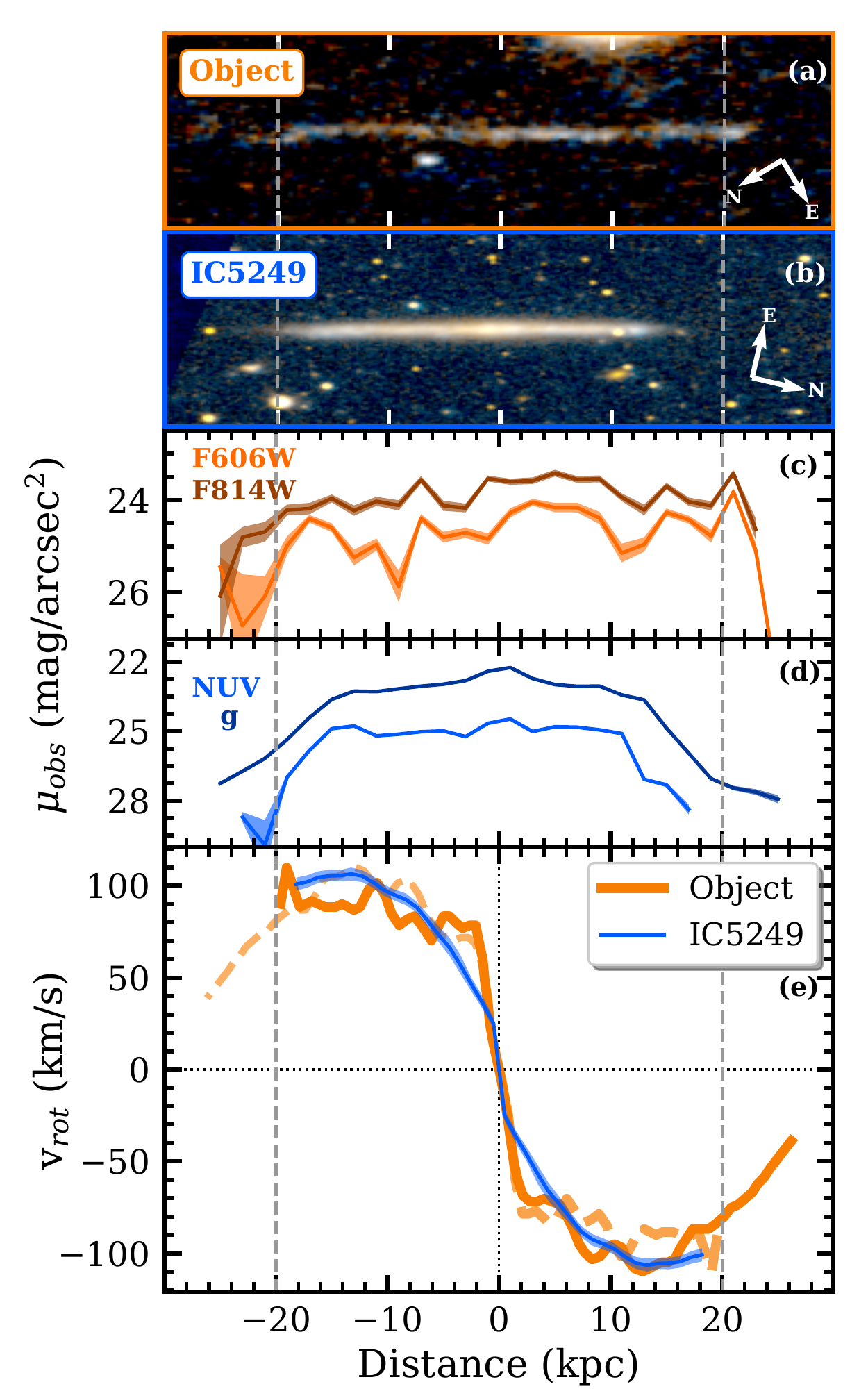}
   \caption{Summary plot with evidence for VD23 object being a bulgeless edge-on galaxy.  
   {\em Panel (a)}: two-color image of the object, i.e., of the conspicuous thin long structure in the image.
   {\em Panel (b)}: two-color image of the local galaxy IC\,5249 presented at the same spatial scale and orientation as the object. The bands (NUV and $g$) were chosen to approximately represent the restframe wavelengths where the object was observed (F606W and F814W).     
   {\em Panel (c)}: Surface brightness profiles of the object in F606W and  F814W as indicated in the inset.
   {\em Panel (d)}: Surface brightness profiles of IC\,5249 in NUV and $g$ as indicated in the inset.
   {\em Panel (e)}: Position--velocity curve of the object from Fig.~10 in VD23 (the thick orange line). The same curve flipped both around $x=0$ (the vertical dotted line) and $y=0$ (the horizontal dotted line) is also represented (the orange dashed lines) to signal the symmetry of the observed position--velocity curve, as expected from the rotation of a galaxy.  The rotation curve of IC\,5249 was also included without any scaling or adjustment \citep[the blue solid line with the light-blue region representing the noise as provided by][]{Banerjee2017}.
   The five panels share the same $x$-axis with the vertical dashed lines, separated by 40\,kpc, included to ease comparison. 
   See Sects.~\ref{sec:data} and \ref{sec:super-thin} for further details.}
\label{fig:allinc}%
\end{figure}
%

\section{Introduction}\label{sec:intro}
Recently, \citet[][hereafter VD23]{2023arXiv230204888V} discovered a remarkable thin extended  object  ($\sim\,$0\farcs 25\,$\times$\,5\arcsec; see Fig.~\ref{fig:allinc}a) roughly aligned with a nearby galaxy. It was  interpreted as the stellar wake induced by the passage of a Super-Massive Black Hole (SMBH) kicked out from the galaxy by the slingshot effect of a 3-body encounter of SMBHs. This object would be the first example of its class as the ejection of SMBHs through this potential mechanism has never been observed so far. It would also provide a new discovery channel for rogue SMBHs and for the process of SMBH ejection from galaxies. In the scenario put forward by VD23, the SMBH producing the wake was ejected with a (projected) speed of $\sim 1600\,{\rm km\,s^{-1}}$ and is observed 39\,Myr after the 3-body encounter, creating a stellar linear structure 40\,kpc long and 2\,kpc wide, with a stellar mass of the order of $3\times 10^9\,{\rm M_\odot}$. As it happens with the ejection, the formation of such a thin  long massive stellar structure may conceptually occur but would be the first observation of this phenomenon. 
Despite the undeniable interest in this eye-catching idea, the actual physical interpretation is not devoid of difficulty.

%

The physical mechanism invoked for the creation of a stellar wake is not specific to SMBHs. It will work for any massive compact object wandering in the halo of a galaxy, including globular clusters, dark matter sub-halos \citep[e.g.,][]{2018PhRvL.120u1101B}, and intermediate-mass black holes \citep[e.g.,][]{2022arXiv221014960D}. Thus, if the physical conditions of the gas in the circum-galactic medium (CGM) that could potentially trigger the formation of a 40\,kpc-long stellar trail were common, then long massive bright stellar trails around galaxies would be ubiquitous. Since this is not the case, the proposed creation of a stellar trail driven by the passage of a SMBH through the CGM of a galaxy is doubly exceptional: it requires the ejection of a SMBH from a galaxy surrounded by a large massive CGM about to self-collapse gravitationally.

Thus, in view of the potential difficulties and to broaden the range of possibilities, here we explore another more traditional explanation of the observation, namely, that the object is a bulgeless galaxy viewed edge-on. Indeed, this idea was mentioned in VD23 but not developed further. 
We use publicly available data (Sect.~\ref{sec:data}) to show that 
the stellar mass of the object plus the observed line-of-sight velocities interpreted as a rotation curve places the object on top of the baryonic Tully-Fisher relation (TRF), that characterizes disk galaxies (Sect.~\ref{sec:super-thin}). 
In addition, we carry out a comparative analysis between the object and the local bulgeless edge-on galaxy IC\,5249 (Sect.~\ref{sec:super-thin};  Fig.~\ref{fig:allinc}b). IC\,5249 has similar properties attributed to the object when interpreted as a bulgeless edge-on galaxy, namely, it has similar stellar mass, rotation curve, disk orientation, extension, and width. 
Evidence for and against the two scenarios (SMBH wake and bulgeless edge-on galaxy) is discussed in Sect.~\ref{sec:discussion}.      
%
Throughout this work, we adopt a cosmological model with  $H_0=70$\,km\,s$^{-1}$\,Mpc$^{-1}$, $\Omega_m=0.3$, and $\Omega_\Lambda=0.7$. All magnitudes are in the AB magnitude system.

\section{Data}\label{sec:data}

We analyze two sources in this paper: the object and the bulgeless galaxy IC\,5249. All used data are archival or from the literature, as declared below.

\subsection{The object}
The object was observed with the Advanced Camera for Surveys (ACS) WideField Channel (WFC) on board of the \emph{Hubble Space Telescope} (\emph{HST}) as part of the program GO-16912 (PI: van Dokkum) available through the MAST archive\footnote{\url{https://mast.stsci.edu/portal/Mashup/Clients/Mast/Portal.html}}. The data consists of one orbit in the F606W band and one in the F814W band, corresponding to an exposure time of 2064\,s in F606W and the same in F814W. 
The charge-transfer efficiency (CTE) corrected data were downloaded from MAST and \emph{flc} files were used to build the final drizzled mosaics with Astrodrizzle \citep{Gonzaga2012}. Figure~\ref{fig:allinc}a shows a color stamp of 7\farcs 5\,$\times$\,3\farcs 5 ($\sim$60 kpc $\times$ 30 kpc at z = $0.964$) centered in the object and generated from F606W and F814W.

Figure \ref{fig:allinc}c shows the surface brightness profiles of the object in the F606W and F814W bands, extracted averaging $\pm1$~kpc along the trace. The shaded areas represent 1-$\sigma$ errors in the profiles, propagated from the photometric error of all the spaxel contributing to each surface brightness bin.

The position--velocity curve shown in Fig.~\ref{fig:allinc}e (the solid orange line) is a replica of the line-of-sight velocity of the object provided by VD23 in the top left panel of their Fig.~10. The velocities in VD23 Fig. 10 between the location of the host galaxy and $\sim$20 kpc are not used for our analysis as they do not represent real measurements but a simple interpolation. The velocities were measured in the [OIII]$\lambda$5007 emission line detected using the LIRIS spectrograph on Keck~I \citep{1995PASP..107..375O}. 
We interpret the position-velocity curve of the object as a rotation curve. In our rendering, $V_{rot}=0$ was estimated as the average between the minimum and maximum velocities of the curve. Then the center of rotation is just the position where the observed $V_{rot}$ is zero. The position thus determined sets the origin of distances in Fig.~\ref{fig:allinc}e, and so in Figs.~\ref{fig:allinc}a and \ref{fig:allinc}c and in the top panel of Fig.~\ref{fig:width}.

According to VD23, the spatially integrated brightness of the object is $\approx40\,\%$ of the brightness of the closest galaxy (the host from which the SMBH was escaping in their proposed scenario). This closest galaxy and the object have  nearly identical colors, implying similar mass-to-light ratios. The stellar mass of the host is estimated in VD23 to be $M_{host}$ $\simeq 7\times10^9\,{\rm M_{\odot}}$, therefore, the inferred mass of the object is $M_\star\simeq 0.4\times M_{host} \simeq 2.8\times10^9\,{\rm M_\odot}$.

\subsection{IC\,5249}

IC\,5249 is a well-known nearby (redshift 0.007) bulgeless  galaxy \citep[e.g.,][]{1999BSAO...47....5K}. In order to compare with the F606W and F814W bands of the object at its redshift, we retrieved the closest available images for this galaxy: NUV \citep[GALEX,][]{Martin2005} and $g$ \citep[DECaLS,][]{decals2016}. Although the $u$-band would be a better choice to replicate F814W, it is not available. 
Figure~\ref{fig:allinc}b shows a color image generated from the NUV and $g$ images. For ease of comparison, it is shown at the same physical scale and orientation as the object. The surface brightness profiles in the two bands are shown in Fig.~\ref{fig:allinc}d, extracted averaging $\pm 1{\rm kpc}$ along the trace of the galaxy. Although barely visible, errors  derived in the same way as for the object are included too. The rotation curve of this galaxy shown in Fig.~\ref{fig:allinc}e (the blue solid line,
with the light-blue region denoting errors) is taken from \citet{Banerjee2017} from HI observations by \citet{Obrien2010c} with the Australia Telescope Compact Array (ATCA).
The distance of this galaxy ($33$\,Mpc) was taken from NED\footnote{\url{https://ned.ipac.caltech.edu/}} to set all physical scales. The value of $V_{max}\simeq 106\,{\rm km\,s^{-1}}$ was obtained directly from the rotation curve. To calculate the stellar mass of IC\,5249, we use the integrated signals in the DECaLS $g$ and $r$ images with the mass-to-light ratio from \citet{Roediger2015} and assuming a \citet{Chabrier2003} initial mass function. The mass of the galaxy thus computed is $\sim 1.9\times 10^9\,{\rm M_\odot}$.

%
\section{Interpretation as a bulgeless edge-on galaxy}\label{sec:super-thin}
\begin{figure}
   \centering
   \includegraphics[width=8cm]{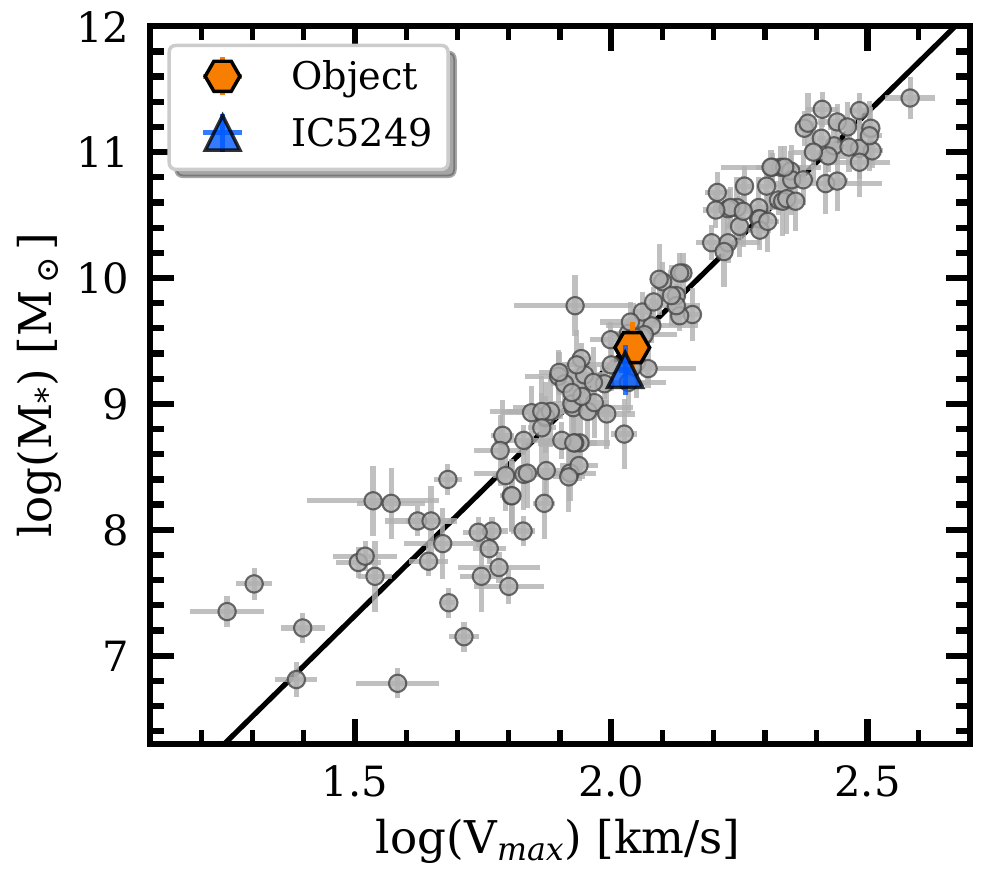}
   \caption{
   Tully-Fisher relation from \citet[][gray points with error bars]{Lelli2019} with the stellar masses from \citet{Lelli2016b}. This rendering represents stellar mass versus maximum rotational velocity. The figure also includes the object (the orange hexagon; $M_\star=2.8\times 10^{9}\,{\rm M_\odot}$, $V_{max}= 110 \,{\rm km\,s^{-1}}$) and the galaxy IC\,5249 (the blue triangle; $M_\star=1.9\times 10^{9}\,{\rm M_\odot}$, $V_{max}= 106 \,{\rm km\,s^{-1}}$). Both objects lie close to each other and right on top of the relation. The solid black line is a linear fit to the grey points of the form $\log M_\star =  A\times\log V_{max} + B$, with $A = 4.0 \pm0.1$ and $B = 1.3\pm0.2$.   }
\label{fig:tullyf}%
\end{figure}
The baryonic TFR \citep[][]{1977A&A....54..661T} links the baryon mass with the maximum rotational velocity of a disk galaxy. In practice, it is a main signpost for being a galaxy, so fundamental as to be routinely employed to determine absolute distances to galaxies from the observed rotational velocities and apparent magnitudes \citep[e.g., the textbooks by][]{2005seg..book.....P,2006asco.book.....H,2006gaun.book.....S}. 
Figure~\ref{fig:tullyf} shows a recent rendering of the stellar TFR as worked out by \citet{Lelli2019} for local galaxies (the stellar masses are from \citealt{Lelli2016b}). Using the stellar mass and the maximum rotational velocity derived in Sect.~\ref{sec:data}, the object happens to appear right on top of the TFR (see the orange hexagon). Repeating the same exercise for IC\,5249 yields very similar results (the blue triangle in Fig.~\ref{fig:tullyf}). 
Thus, the location of the object on the TFR provides solid support to our conjecture that it is a galaxy.  

More evidence can be extracted from Fig.~\ref{fig:allinc}. The images of the object and the local galaxy are quite similar both in extension and width (cf. the two top panels of the figure). The image of the object is clumpier or more irregular, but this is to be expected because (1) high redshift galaxies are in general more actively forming stars and therefore clumpier \citep[e.g.,][]{2005ApJ...627..632E,2005ApJ...631...85E}, and (2) the redder band used for the object (F814W, corresponding to restframe wavelengths between 3500 and 4700\,\AA) is significantly bluer than the $g$ color filter (from 4000 to 5500\,\AA) used for IC\,5249. It is well known that star-forming galaxies appear more irregular in bluer bandpasses \citep[e.g.,][]{2001AJ....122..729K,2018ApJ...864..123M}.

The surface brightness profiles of IC\,5249 and the object along the semi-major axes are also quite similar (Figs.~\ref{fig:allinc}c and \ref{fig:allinc}d). These surface brightness profiles were obtained using an aperture of 2 kpc for both objects. Both profiles are fairly flat with a sharp drop in the outskirts. The radial position of the edges \citep[where the sharp drops are located; see][]{2020MNRAS.493...87T,2022A&A...667A..87C} are R$_{edge}$=15\,kpc for IC\,5249 and 20\,kpc for the object. The edges of both objects tend to be bluer (Figs.~\ref{fig:allinc}c and \ref{fig:allinc}d). There are some differences in the global colors (IC\,5249 is redder) and the presence of a small bulge in IC\,5249 which is not visible in the object.    

The position--velocity curve of the object provided by VD23 (their Fig.~10, top left panel)  is plotted as a solid orange line in Fig.~\ref{fig:allinc}e.  The same curve flipped both around $x=0$  and $y=0$ is also represented (the orange dashed lines) to highlight the anti-symmetry of the observed position--velocity curve, as expected from the rotation of a galaxy. The blue solid line represents the rotation curve of IC\,5249 from \citet[][]{Banerjee2017}, and it was included without any scaling or adjustment. The agreement between the object and  IC\,5249 is remarkable, supporting that the position--velocity curve of the object is indeed the rotation curve of a galaxy. In addition to the overall agreement in shape and amplitude, we note the tendency of both curves to drop in the outskirts, a property often observed in high redshift galaxies \citep[e.g.,][]{2017Natur.543..397G} that tells us about the relative importance of baryons to dark matter.

\begin{figure}
   \centering
   \includegraphics[width=8cm]{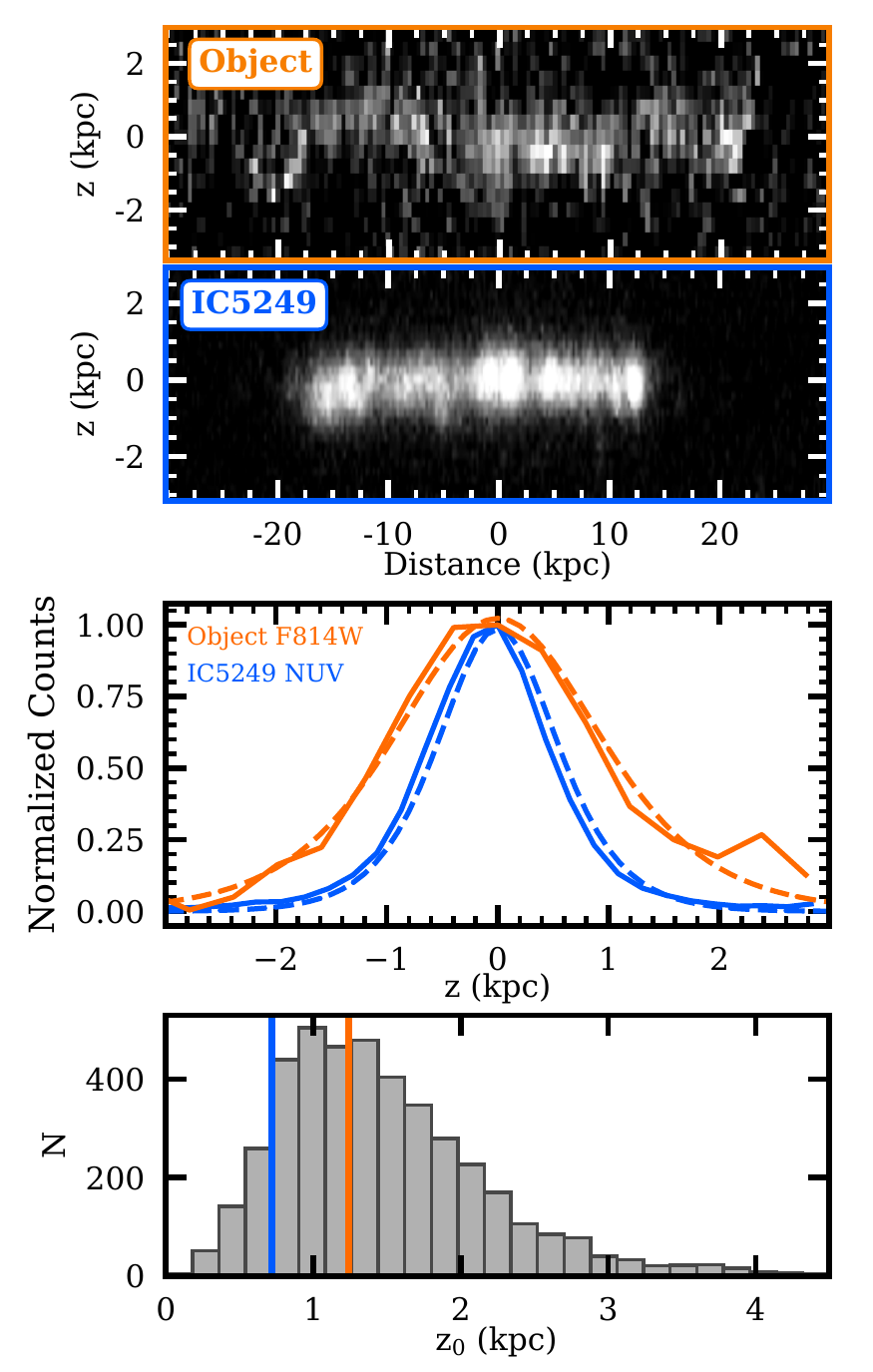}
   \caption{Top panels: F814W image of the object and NUV image of IC5249. Middle panel: Normalized profiles in counts along the vertical direction (along $z$ in the top panels). The dashed lines show ${\rm sech}^2(z/z_0)$ fits to these vertical cuts (see a detailed description in Sect.~\ref{sec:super-thin}). Bottom panel: Histogram of scale-heights for edge-on galaxies from \citet{2014ApJ...787...24B}. The vertical colored lines indicate the values of the scale-heights from the fit for the object (orange) and IC\,5249 (blue).
   }
\label{fig:width}%
\end{figure}

We have also investigated whether the width of the object is compatible with being a bulgeless galaxy. Figure~\ref{fig:width} shows images of both the object (in the F814W band) and IC\,5249 (in the NUV band). Visually, both have similar widths. To quantify this statement, the central panel of Fig.~\ref{fig:width} shows the distribution of counts across the object and IC\,5249. To create these profiles, the distribution of counts of both galaxies was collapsed along the long (horizontal in Fig. \ref{fig:width}) axes. By fitting the vertical distribution of counts to the commonly used function ${\rm sech}^2(z/z_0)$, with z$_0$ the scale-height, we obtain $z_0$=0.72\,kpc for  IC\,5249 \citep[in very nice agreement with the value of 0.724\,kpc reported by][]{Banerjee2017} and $z_0$=1.24\,kpc for the object. The fits are shown as dashed lines in the central panel of Fig.~\ref{fig:width}. The bottom panel of the same figure compares the derived values of $z_0$ with the  scale-heights of the sample of edge-on disks from \citet{2014ApJ...787...24B}. The width of both IC\,5249 and the object are perfectly normal in terms of scale-height, a result well known for bulgeless galaxies \citep[see, e.g.,][]{2017MNRAS.465.3784B}.

%
\section{Discussion}\label{sec:discussion}

The main arguments for the object to be a bulgeless edge-on galaxy are: (1) its position--velocity curve resembles a rotation curve which, together with its stellar mass, puts the object right on top of the TFR (Fig.~\ref{fig:tullyf}), and (2) the rotation curve, stellar mass, extension, width, and surface brightness profile of the object is very much like IC\,5249 (Fig.~\ref{fig:allinc}), which without much forcing could be a local analog.   

However, there are three additional observables mentioned by VD23 that require further analysis to be accommodated into the edge-on disk scenario. First, the object is not completely straight (Fig.~\ref{fig:allinc}, top panel), with variations of $<$\,0.5\,kpc. This could be due to warps in the disk or simply because we are observing in the UV restframe, where galaxies are more irregular since the UV traces individual star-forming regions (see the discussion in Sect.~\ref{sec:super-thin}). 
Second, the line ratio [OIII]/H$\beta$ exceeds the canonical limit for HII regions in two positions along the stellar trace, reaching the value of 10 expected for shocks or AGNs. However, the observed H$\beta$ line is extremely weak and noisy in these points, and a factor two uncertainty in its flux would bring the ratio back to HII region values. Moreover, some extreme star-forming galaxies \citep[e.g., green peas and the like;][]{2014ApJ...795..165S,2020Galax...8...13L} provide ratios very close to the largest found in the object. Another possibility is that the emission is tracing shocks created by star formation in HII regions, without the intervention of any SMBH. Those shocks are not uncommon when SN-driven bubbles meet the interstellar medium (ISM) or can be caused by the random motions associated with the ISM turbulence \citep[e.g.,][]{2014A&A...563A..49S,2022A&A...660A..77D}.  
Third, the change of color along the stellar trace has a pattern that reproduces the evolution of a young single stellar population with time (Fig.~9 in VD23), indicating linearly increasing aging from left to right in Figs.~\ref{fig:allinc}a and \ref{fig:allinc}c.  Within the stellar disk scenario, the changes in color would have to be produced by the random mixture of younger and older populations that co-exist in the disk. This explanation avoids a serious issue of the systematic age variation explanation of colors, namely, the absence of a systematic surface brightness variation of the stellar trace (Fig.~\ref{fig:allinc}c). With some dependence on the details of the single-stellar population modeling and on the bandpass of observation, we should expect a systematic luminosity variation from the oldest stars (leftmost point in Fig.~\ref{fig:allinc}, panel a) to the youngest stars (rightmost point) of around 2 magnitudes \citep[e.g.,][]{1999ApJS..123....3L,2009ApJ...699..486C}. This significant variation in magnitude is simply not observed. 

The overdensity expected from the passage of a SMBH is not large which implies very special physical conditions in the IGM for the star formation to be triggered.
If such conditions were regularly met it would imply that the passage of any massive compact object, like a globular cluster, will trigger star formation, and galaxies will be plagued  with conspicuous stellar wakes. Thus, the scenario where the object is produced by the ejection of a SMBH is doubly exceptional: it requires the ejection of a SMBH from a galaxy that then encounters a huge massive gas cloud about to self-collapse gravitationally.
In addition, the SMBH scenario does not explain the two main observational features supporting the galaxy hypothesis, namely, why the object fits in the TFR (Fig.~\ref{fig:tullyf}) and why its main observational properties agree so well with those of a local bulgeless galaxy with similar stellar mass (IC\,5249; Fig.~\ref{fig:allinc}).   
Other observables where the SMBH wake scenario seems to do a remarkably good job actually face problems too. These observables are the color variation of the trail and its transverse displacement according to time from the SMBH pass. If the change of color would be produced by the aging of the stellar population, the surface brightness should change systematically along the stellar trace, being some two magnitudes fainter in the older end (see the discussion above). As for the transverse displacement, the agreement relies on the implicit assumption that the gas cloud transverse  velocity is zero in the older end of the stellar trace. Even if this might happen, it represents an ad-hoc assumption without which the  agreement between the observed transverse displacement and line-of-sight velocity goes away.

In short, in this Letter we present a number of solid arguments supporting the scenario where the object discovered by VD23 could be a bulgeless edge-on galaxy. If it is, it would not be a common galaxy in the sense of being bulgeless and particularly long for its stellar mass and redshift. However, it is not very uncommon either since thousands of this kind of galaxies (often called thin galaxies or flat galaxies) are already known \citep[e.g.,][]{1999BSAO...47....5K}. One of them, IC\,5249, has been used here to highlight the close resemblance of the object with these galaxies.

\begin{acknowledgements}
  Thanks are due to Mar\'\i a Benito for discussions on the formation of a stellar wake.
Thanks are also due to an anonymous referee for useful suggestions to improve the clarity of our arguments.
JSA acknowledges financial support from the Spanish Ministry of Science and Innovation, project PID2019-107408GB-C43 (ESTALLIDOS), and from Gobierno de Canarias through EU FEDER funding, project PID2020010050.
MM acknowledges support from the Project PCI2021-122072-2B, financed by MICIN/AEI/10.13039/501100011033, and the European Union “NextGenerationEU”/RTRP and IAC project P/302302.
IT acknowledges support from the ACIISI, Consejer\'{i}a de Econom\'{i}a, Conocimiento y Empleo del Gobierno de Canarias and the European Regional Development Fund (ERDF) under grant with reference PROID2021010044 and from the State Research Agency (AEI-MCINN) of the Spanish Ministry of Science and Innovation under the grant PID2019-107427GB-C32 and IAC project P/302300, financed by the Ministry of Science and Innovation, through the State Budget and by the Canary Islands Department of Economy, Knowledge and Employment, through the Regional Budget of the Autonomous Community.
\end{acknowledgements}

%

\end{document}